\documentclass[12pt]{article}
\usepackage{cjp}
\usepackage{epsf}

\begin{document}

\title{
Techniques for using the overlap-Dirac operator to
                             calculate hadron spectroscopy.\footnote{
Talk presented at Chiral '99, Sept 13-18, 1999, Taipei, Taiwan.}
}

\author{UKQCD Collaboration: Craig~McNeile.
\address{Department of Mathematical Sciences, 
University of\ Liverpool, L69 3BX, UK}
%  end of author list
}

\maketitle
 
\begin{abstract}
We report on our progress in using the overlap-Dirac fermion operator
in simulations of lattice QCD. We have investigated the Lanczos based
method of Borici, as well as various rational approximations, to
calculate the step function in the overlap-Dirac operator.  The QCD
simulations were performed at $\beta = 5.85$ with a lattice volume of
$8^3 \; 32$.

\end{abstract}
\section{INTRODUCTION}

The results of lattice QCD calculations of weak matrix elements are a
critical component of the experimental program in heavy flavour and
kaon physics.  The results from lattice gauge theory calculations
would significantly improve if the masses of both the sea and valence
quarks could be reduced.  Unfortunately progress in doing this is very
slow with simulations that use either the Wilson or clover fermion
operators~\cite{Gottlieb:1997hy}.

It seems plausible that the difficulty of simulating with light quark
masses with the clover and Wilson fermions operators is due to
explicit chiral symmetry breaking in the actions. If the fermion operators
were invariant under chiral symmetry transformations, their eigenvalue
spectrum would be constrained
to a smaller region~\cite{Neuberger:1999zk}. The performance of the
simulation algorithms degrades as the range of eigenvalues gets
larger.  Simulations that use the staggered fermion operator can reach
much lower quark masses~\cite{Gottlieb:1997hy} than simulations that
use the Wilson or clover operators, because the staggered action has a
residual of the continuum chiral symmetry.  Neuberger has
derived~\cite{Neuberger:1997fp} a fermion operator, called the
overlap-Dirac operator, that has a lattice chiral
symmetry~\cite{Ginsparg:1982bj,Luscher:1998pq}.

Our goal is to simulate QCD with the overlap-Dirac operator in the
mass region: ($M_{PS}/M_V = 0.3 - 0.5$).  This quark mass region is
inaccessible to lattice QCD simulations, that are currently
computationally feasible, with the clover or Wilson
operators~\cite{Gottlieb:1997hy}.

\section{THE OVERLAP-DIRAC OPERATOR}

The massive overlap-Dirac
operator~\cite{Neuberger:1997fp,Edwards:1998wx} 
is 
\begin{equation}
D^{N} = \frac{1}{2}( 1 + \mu +
(1-\mu) \gamma_{5} \frac{H(m)}{\sqrt{H(m)H(m) }}     )
\end{equation}%%
where $H(m)$ is the hermitian Wilson fermion operator with negative
mass, defined by
%%%
\begin{equation}
H(m) = \gamma_5 ( D^W - m)
\label{eq:gfiveWilson}
\end{equation}
%%%
%%%
where $D^W$ is the standard Wilson fermion operator.
The parameter $\mu$ 
is related to the physical quark mass and lies in the range
$0$ to $1$.
%%%
The $m$ parameter is a regulating mass, in the range between a
critical value and 2.
%%%
%%%

\section{NUMERICAL TECHNIQUES}

Quark propagators are calculated using a sparse matrix inversion
algorithm.  The inner step of the inverter is the application of the
fermion matrix to a vector. For computations that use the 
overlap-Dirac,
the step function
\begin{equation}
\epsilon(H) \underline{b} = 
\frac{H} { \sqrt{H  H } }  \underline{b}
\label{eq:stepfunc}
\end{equation}
must be computed using some sparse matrix algorithm.  A number of
algorithms that calculate quark propagators from the overlap-Dirac
operator, without using a nested algorithm have been
proposed~\cite{Neuberger:1999re,DeGrand:1999xp}.  We discuss one of
them in section~\ref{se:fiveD}.

Practical calculations of the overlap operator are necessarily approximate.  To
judge the accuracy of our approximate calculation we used the
(GW) Ginsparg-Wilson error:
\begin{equation}
\mid \mid 
 \gamma_5 D^N \underline{x} 
+   D^N  \gamma_5  \underline{x} 
   - 
  2 D^N  \gamma_5 D^N  \underline{x} 
\mid \mid 
\frac{1}{\mid\mid \underline{x} \mid\mid  }
\label{eq:GWerrDEFN}
\end{equation}
%%
%%  technical definition of the norm
%%
%%
which just checks that the matrix obeys the 
Ginsparg-Wilson relation~\cite{Ginsparg:1982bj}.
Other groups~\cite{Edwards:1999mp,Liu:1999aa}
 have found that the step function 
must be calculated very accurately, so we also 
use more sophisticated estimates of the numerical
error (see Eq.~\ref{eq:GOM}).

%%
%%  numerical parameters
%%

Our numerical simulations were done using $\beta = 5.85$ quenched
gauge configurations, with a volume of $8^3 \; 32$.  This allows us to
directly compare our results with the two other QCD spectroscopy
calculations~\cite{Edwards:1999mp,Liu:1999aa}.  The quark propagators
were generated from point sources. For all the algorithms we
investigated, we used $m$ equal to $1.5$.  Although we are currently
investigating using the overlap-Dirac operator in the quenched theory,
most of the algorithms can also be used in
full QCD simulations~\cite{Kennedy:1998cu,Liu:1998hj}.

\section{LANCZOS BASED METHOD}

Borici~\cite{Borici:1998mr} has developed a method 
to calculate the action of the overlap-Dirac operator on a vector, using
the Lanczos algorithm.
%%
%%  references
In exact arithmetic,
the Lanczos algorithm generates an orthonormal
set of vectors that tridiagonalises the matrix.
%%%
\begin{equation}
H Q_n = Q_n T_n 
\end{equation}
%%%
where $T_n$ is a tridiagonal matrix. The columns of $Q_n$ contain
the Lanczos vectors.

The ``trick'', to evaluate the step function (Eq.~\ref{eq:stepfunc}),
is to set the target 
vector $\underline{b}$, as the first vector in the Lanczos 
sequence.
%%%
An arbitrary function $f$ of the matrix $H$ 
acting on a vector is constructed using
%%%
\begin{eqnarray}
(f(H) b )_i & = & \sum_{j} ( Q_n f(T_n) Q_n^\dagger )_{i\; j} b_j \\
& = & \| b \|  ( Q_n f(T_n)  )_{i\; 1} \label{eq:what_to_code}
\end{eqnarray}
%%%
where the orthogonality of the Lanczos vectors has been used.  The
$f(T_n)$ matrix is computed using standard dense linear algebra
routines. For the step function the eigenvalues of $T_n$ are replaced
by their moduli.  Eq.~\ref{eq:what_to_code} is linear in the Lanczos
vectors and thus can be computed in two passes.
%%%
%%%

The major problem with the Lanczos procedure is the loss of the
orthogonality of the sequence of vectors due to rounding errors. It is
not clear how this lack of orthogonality effects the final results.
Some theoretical analysis has been done on the use of the 
Lanczos algorithm to calculate functions of 
matrices~\cite{Druskin:1998:UNL}. It is claimed that the lack of
orthogonality is not important for some classes of functions.

On small lattice we checked that the eigenvalue spectrum of the
overlap-Dirac operator moves closer to a circle~\cite{McNeile:1999zy},
as the number of Lanczos steps increases.  Even after 50 iterations of
the Lanczos algorithm, there are still small deviations from the
circle.  For a hot gauge configuration 
with a volume of  $4^4$, all the Lanczos vectors were stored
and then were used to investigate the effect of the loss of orthogonality.
The scalar product between two Lanczos vectors drops from $10^{-7}$ to
$10^{-3}$ after about 130 iterations, indicating problems with
orthogonality. The GW error was $0.11$ at 50 iterations and $0.012$ at
250 iterations. This is some evidence that the Borici's algorithm may
still work even when the orthogonality of the Lanczos vectors is lost.
%%%
%%%
%%%
%%%
It is much harder to look at the eigenvalue spectrum of the
overlap-Dirac operator on a $8^3 \; 32$ $\beta=5.85$ gauge
configuration, so we computed the GW error instead. The GW error 
on a
single gauge configuration 
was:
$2\; 10^{-3}$ (100 iterations), $1\; 10^{-3}$ (200) iterations, $1\;
10^{-4}$ (300) iterations, and $2\; 10^{-6}$ (500 iterations).

\begin{figure}
\vbox{\epsfxsize=2.7in \epsfbox{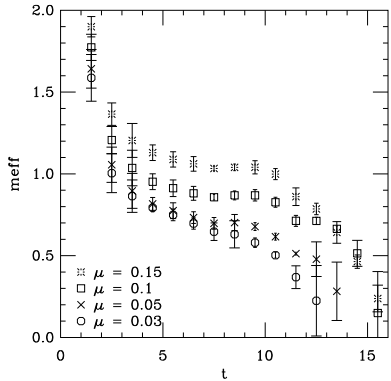}}
%%\vskip -0.5in
%%\vskip -9mm
\figcaption{Effective mass plot for pion using Borici's method of
calculating the step function using 100 Lanczos iterations.}
\label{fig:meff}
%%\vskip -8mm
\end{figure}

Fig.~\ref{fig:meff}, is an effective mass plot of the pion, for four
different values of the quark mass $\mu$. The number of iterations in
the Lanczos algorithm was kept constant at $100$.  The effective mass
plots for the pion in Fig.~\ref{fig:meff} can be compared to the two
other published spectroscopy calculations, that both use
configurations with parameters: $8^3 16$ and $\beta = 5.85$.  For a
quark mass of $\mu = $ $0.1$ ($0.05$), Liu et al.~\cite{Liu:1999aa}
obtain a pion mass in lattice units of $0.63$ ($0.45$).  From the
graph by Edwards et al.~\cite{Edwards:1999mp}, the pion mass was
$0.87$ ($0.63$) at $\mu$ = $0.1$ ($0.05$).
The differences between the two groups are probably explained by 
them using different values of $m$ in their simulations, because
the
quark mass $\mu$ is renormalized by a multiplicative factor that
depends on the domain mass~\cite{Edwards:1998wx}.  The effective mass
plots in Fig.~\ref{fig:meff} are consistent with the data of Edwards
et al.~\cite{Edwards:1999mp}.  Although smeared correlators should be
used for a more detailed comparison.  The quality of the $\mu$ = 0.03
effective mass plot of the pion is disappointing. The inversion of the
overlap-Dirac operator that used 100 Lanczos iterations, at a mass
$\mu = 0.1$, required 150 iterations in the inverter and took 105
minutes on 32 nodes of our cray t3e.

%%
%%  compare with Edwards/Liu
%%
We checked the stability of the pion's effective mass 
with the number of iterations used in the Lanczos
procedure. The pion effective mass plot was stable 
for the quark masses $\mu=$  $0.1$ and $0.03$, as the 
number of Lanczos iterations were varied from 100 
to 300.

Liu et al.~\cite{Liu:1999aa} used the Gell-Mann-Oakes-Renner (GOR) relation,
that was derived in~\cite{Edwards:1998wx} for the overlap-Dirac
operator, as a check on the accuracy of the computation of the step
function. 
\begin{equation}
\mu \sum_x \langle \pi(x) \pi(0) \rangle = 
\frac{1}{V}
\sum_x \langle \overline{\psi}(x)  \psi(x)  \rangle
\label{eq:GOM}
\end{equation}
where $\pi$ is the pion interpolation field,
and $x$ is summed over the space-time volume (V).
The ``external'' quark propagators~\cite{Edwards:1998wx}
defined by
\begin{equation}
\hat{D}(\mu) = \frac{1}{1 - \mu} [ D^{-1}(\mu) -1 ]
\end{equation}
should be used in equation~\ref{eq:GOM}.

\begin{figure}
\vbox{\epsfxsize=2.7in \epsfbox{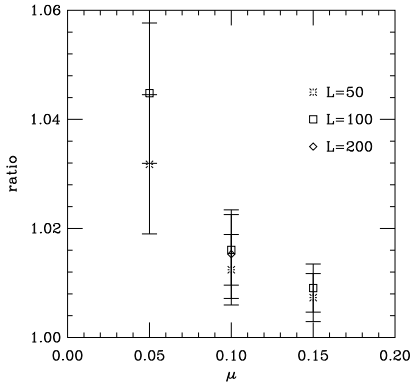}}
%%\vskip -0.5in
%%\vskip -9mm
\figcaption{
The ratio of the right hand side over the left hand side 
in equation\protect{~\ref{eq:GOM}} for a single configuration.
The error bars are from the $Z_2$ noise method used to compute
the chiral condensate. The step function was computed
using Borici's method with: $50$, $100$, $200$ iterations.
}
\label{fig:GMOcheck}
%%\vskip -8mm
\end{figure}

The data in Fig.~\ref{fig:GMOcheck} show the GOR relation is satisfied
up to 2\% for the masses $\mu = $ $0.1$ and $0.15$, and 4\% for the
mass $\mu$ = $0.05$. Increasing the number of Lanczos iterations did
not decrease the violation of the GOR relation. This may be due to the
loss of orthogonality in the Lanczos vectors.

\section{RATIONAL APPROXIMATION}
The step function can be approximated by a rational
approximation~\cite{high94p,Neuberger:1998my}.
\begin{equation}
\epsilon(H) \sim H \; ( c_0 + \sum_{k=1}^{N} 
   \frac{c_k}{H^2 + d_k}  )
\label{eq:pole}
\end{equation}
The rational approximation is only an accurate approximation to the
step function in a certain region.  The eigenvalues of the matrix $H$
should lie in this region. The coefficients $c_k$ and $d_k$ can be
obtained from the Remez algorithm~\cite{Edwards:1998yw}.  The number
of iterations required in the inverter is controlled by the smallest
$d_k$ coefficient, that acts like a mass.  If $d_k$ is small, the
number of iterations required is controlled by the condition number of
$H^{2}$.

On one configuration we obtained GW errors of: $8\; 10^{-5}$, $1\;
10^{-5}$, and $2 \; 10^{-6}$, for the $N=6$, $N=8$, and $N=10$,
optimal rational approximations~\cite{Edwards:1998yw}.  Unfortunately,
the above results required up to 600 iterations for the smallest
$d_k$, which was too large to use as the inner step of a quark
propagator inverter.  We have not yet implemented the technique of
projecting out some of the low lying eigenmodes~\cite{Edwards:1998yw}.
This projection will reduce the condition number of the matrix, and
hence the number of iterations required in the inner inversions.

\section{Five dimensional representation of the overlap-Dirac operator}\label{se:fiveD}

One undesirable feature of the algorithms just presented for inverting
the overlap-Dirac operator is that at each iteration of the inverter,
some sparse matrix techniques must be done to calculate the
overlap-Dirac operator. In the language of Krylov spaces, the theory
of which underlies the numerical calculations, two independent Krylov
spaces are used in a nested inverter.  If the overlap-Dirac operator
could be calculated using one Krylov space, this may be more
efficient.  Neuberger has proposed
one method to calculate the overlap-Dirac operator without the nested
inversion~\cite{Neuberger:1999re}.  The first implementation of
Neubeger's ideas was discussed by Edwards at this meeting.

To explain the idea we will use a simplified rational
approximation. The generalisation to higher order rational
approximations is obvious.
\begin{equation}
\left(\frac{1}{2} (1 + \mu)  + \frac{1}{2} (1 - \mu)  \gamma_5 c_0 H 
\frac{(H^2 + p_1)(H^2 + p_2) }{(H^2 + q_1)(H^2 + q_2)}
\right)
\psi = b
\end{equation}
%%%
The equation for $\psi$ can be solved using 
additional variables ($\phi_{i}$). The additional equations 
generated by the new variables can be written in matrix form.
%%
%%%\[
\begin{equation}
\left(
\begin{array}{ccccc}
    -1 &     0      &  0 & 0 & H_{p1}   \\
    -1 &  H_{q1}  &  0 & 0 &  0          \\
     0 &  H_{p1}  & -1 & 0 &  0          \\
     0 &   0      & -1 & H_{q1} &  0      \\
  0 &  0 & 0 & \frac{1}{2}(1 -\mu) \gamma_5 c_0 H  & \frac{1}{2}(1 +\mu) 
\end{array}
\right)
\left(
\begin{array}{c}
     \phi_1 \\
     \phi_2 \\
     \phi_3 \\
     \phi_4 \\
     \psi \\
\end{array}
\right)
=
\left(
\begin{array}{c}
     0 \\
     0 \\
     0 \\
     0 \\
     b \\
\end{array}
\right)
\label{eq:5Dmatrix}
%%\]
\end{equation}
where we have introduced the notation
$H_c = H^2 + c$.

The additional variables makes the calculation of the overlap-Dirac
operator look similar to the calculation of the domain wall
operator~\cite{Blum:1998ud}.  Although with an accurate enough
rational approximation, this technique will calculate the
overlap-Dirac operator exactly.  The key issue is the condition number
of the five dimensional matrix, because this controls the number of
iterations required in the inverter. As the various rational
approximations use small coefficients, these could have a large effect
on the condition number.

To study the effect of the rational approximation on the condition
number of the five dimensional matrix, we have started to study the
problem in free field theory. The calculation of the eigenvalues of
the matrix in Eq.~\ref{eq:5Dmatrix} is simple in free field, because
Fourier analysis can be used.  Although the free field theory
eigenvalues will not be too similar to those of the interacting theory
(although projecting out the lowest topological eigenmodes will
improve the agreement), this is the only case where we have any hope
of analytical insight into the condition number of the five
dimensional matrix.

The five dimensional matrix was constructed
using MATLAB, using the free hermitian Wilson
operator~\cite{Carpenter:1985dd}.  Table~\ref{eq:freeRE} contains some
results for a $8^4$ lattice, with a quark mass of $\mu = 0.1$, and a
domain mass of $m = 1.0$. The $N16$ approximation is the $16$th order
approximation to the step function introduced by Neuberger and
Higham~\cite{high94p,Neuberger:1998my}. The $R6$, $R8$, and $R10$ rows
are for the Remez approximations to the step function introduced by
Edwards et al.~\cite{Edwards:1998yw}.  The validity column is the
maximum distance that the rational approximation deviates by $10^{-3}$
from unity, divided by the minimum distance. This is a measure of how
good an approximation the rational function is. The order in
Table~\ref{eq:freeRE} has been normalised so that it is comparable to
the length of the lattice in the fifth dimension for domain wall
fermions (the true order is obtained by multiplying by 12).

Table~\ref{eq:freeRE} shows that the condition number
of the five dimensional matrix strongly depends on the type of rational
approximation used to construct it. 
It is interesting to compare the 
$N16$ and $R8$ approximations that are almost equally
good, but which have very different condition numbers.
It would be instructive to compare the condition numbers 
in Table~\ref{eq:freeRE} with the condition number of 
the domain wall fermion operator~\cite{Blum:1999xi}.

\begin{table}
\begin{tabular}{ccc|c}
Approximation & Validity & Order & Condition number \\ \hline
N16  &  $69$  & $32$ & $790$ \\
R6   &  $16$  & $13$  & $1460$ \\
R8   &  $61$  & $17$  & $2580$ \\
R10  &  $303$ & $21$  & $4260$ \\
\end{tabular}
\tblcaption{Condition number of the five dimensional matrix}
\label{eq:freeRE}
\end{table}

\section{CONCLUSIONS}

The results for the masses of the light hadrons obtained by Liu et
al.~\cite{Liu:1999aa} and Edwards et al.~\cite{Edwards:1999mp} seem to
show that QCD simulations can be run at much lighter quark masses,
than can be explored with the standard Wilson or clover operators.  To
work with quarks as light as those in the calculations of Edwards et
al.  and Liu et al., we need to project out the lowest eigenvalues
from the $H$ matrix. We are working on an implementation of the
eigenvalue projection technique.

What is not clear is how expensive the simulations with the
overlap-Dirac operator will be on more realistic lattice volumes.  The
only way the Wilson or clover operators can be used to simulate QCD
with lighter quarks is by the brute force approach of simulating
closer to the continuum limit. This too will be very expensive.
As the overlap-Dirac operator has a  lattice chiral symmetry,
it should be able to be used to explore the light quark
mass region of QCD in an elegant way.
%
%  Acknowledgements
%

This work is supported by PPARC.  The computations were carried out on
the T3E at EPCC in Edinburgh.  We thank K. Liu, T. Kennedy,
R. Edwards, C. Michael, A. Irving and U. Heller for discussions.

%%
%%

%%%\bibliographystyle{prstytad}
%%\bibliographystyle{h-physrev2}
%%\bibliography{conferences,biblio,overlap}

\end{document}